\documentstyle[12pt,psfig] {article}

\newcommand{\be}{\begin{equation}}
\newcommand{\ee}{\end{equation}}
\newcommand{\bd}{\begin{displaymath}}
\newcommand{\ed}{\end{displaymath}}
\newcommand{\ba}{\begin{array}}
\newcommand{\ea}{\end{array}}
\newcommand{\bt}{\begin{tabular}}
\newcommand{\et}{\end{tabular}}
\newcommand{\bc}{\begin{center}}
\newcommand{\ec}{\end{center}}
\newcommand{\bn}{\begin{enumerate}}
\newcommand{\en}{\end{enumerate}}
\newcommand{\bi}{\begin{itemize}}
\newcommand{\ei}{\end{itemize}}
\newcommand{\bqr}{\begin{eqnarray}}
\newcommand{\eqr}{\end{eqnarray}}
\newcommand{\bfig}{\begin{figure}[tbp]}
\newcommand{\efig}{\end{figure}}
\newcommand{\btab}{\begin{table}[ht]}
\newcommand{\etab}{\end{tabular}\ec\end{table}}
\newcommand{\bl}{\begin{large}}
\newcommand{\el}{\end{large}}
\newcommand{\vspc}{\vspace{0.3cm} \bc}

\newcommand{\hh}{\hline\hline}
\newcommand{\nb}{\nonumber}

\newcommand{\ket}[1]{\vert #1 \rangle}

\newcommand{\nuc}[2]{\mbox{\relax\ifmmode{}^{#1}{\protect\text{#2}}\else${}^{#1}$#2\fi}}
\title      {Pairing correlations. \\
            Part 1: description of odd nuclei in mean-field theories.}
\author     {
             T. Duguet, P. Bonche, \\
             {\em Service de Physique Th\'eorique, CEA Saclay,} \\
             {\em 91191 Gif sur Yvette Cedex, France} \\
	     P.-H. Heenen, \\
	     {\em Service de Physique Nucl\'eaire Th\'eorique, Universit\'e Libre} \\ 
	     {\em de Bruxelles, C.P 229, B-1050 Bruxelles, Belgium} \\
\and	     J. Meyer \\
             {\em Institut de Physique Nucl\'eaire de Lyon, } \\
	     {\em CNRS-IN2P3 / Universit\'e Claude Bernard Lyon 1,} \\
	     {\em 43, Bd. du 11.11.18, 69622 Villeurbanne Cedex, France}
            }

\begin{document}
 
\maketitle
 
\begin{abstract}
In order to extract  informations on pairing correlations in nuclei 
from experimental masses, the different contributions to odd-even mass 
differences are investigated within the Skyrme HFB method. In this first paper, 
the description of odd nuclei within HFB  is discussed since it is 
the key point for the understanding of the above mentioned contributions. 
To go from an even nucleus to an odd one, the advantage of a two steps process is 
demonstrated and its physical content is discussed. New results concerning time-reversal symmetry breaking in odd-nuclei are also reported.



\end{abstract}

\footnote{Corresponding author~: duguet@spht.saclay.cea.fr}


\section{Introduction}
\label{intro}

A proper description of odd nuclei by mean-field methods 
requires to break the time-reversal symmetry, making their study much harder 
than for even ones. Since this symmetry is broken by the unpaired nucleon, 
the BCS approximation is not anymore
valid and has to be replaced by the Hartree-Fock-Bogolyubov (HFB) one.
 This symmetry breaking has also the consequence that the
individual wave-functions are no longer doubly
degenerate, doubling at least the computing task.
Nevertheless, because of present computer capacities and of the development of new
iteration schemes, it is now possible to describe 
even and odd nuclei on the same footing at the mean-field level of 
approximation.

Thanks to that, observables  can be calculated
along an isotopic or isotonic chain without uncertainties
related to a different level of approximations for 
even and odd particle number.
This is particularly important for
differential quantities computed by finite difference formulae, as the 
odd-even mass staggering (OES). Such observables, directly related
to experimental data, 
put into evidence the specificities of odd nuclei
with respect to even ones and have been intensively used to adjust
effective pairing interactions.   

Their proper analysis is difficult as self-consistency
mixes the different effects related to the addition of a nucleon,
especially the modification of the chemical potential, the breaking of
time-reversal symmetry and the weakening of pairing correlations. 
In order to isolate an interpret the different contributions to odd-even effects, it is 
essential to correctly formalize and understand the transition 
between even and odd quantum states. To give some insights on this question, 
a perturbative analysis is particularly adapted since one can a priori write 
analytical relations between neighbor nuclei in term of creation or annihilation 
operators.

The HF approximation provides a useful step in the understanding
of this transition, since it does not involve pairing effects. 
In this case, the
link~\cite{ring}  between even and odd states  is perturbatively given by the
creation of a particle on the first empty level in the state
of the nucleus with one less nucleon. 
We shall however reconsider this simple case because, to be useful for
the understanding of the more general HFB description of odd nuclei,
the HF approximation has to be derived as the zero pairing limit of
the HFB one. 

When pairing is taken into account, a well-known successful perturbative 
procedure consists in describing an odd state as a one 
quasi-particle (qp) state on the even neighbor vacuum. 
However, this procedure suffers from an inconsistency with regard to the particle 
number~\cite{ring}. It demands an ad hoc readjustment of the chemical potential. 
This is what is implicitly done when the 
theoretical BCS gap at the Fermi energy taken from the calculation of an even 
state is compared to experimental odd-even mass differences. This procedure is 
however quite satisfactory for energy predictions, and has been 
extensively used~\cite{garrett}.

We shall show that these inconsistencies can eliminated 
by a redefinition of the vacuum on which
the qp is created by perturbation.
The two step prodedure which is introduced allows to 
analyse in details the description of odd nuclei by fully 
self-consistent calculations and
in particular, to emphasize the changes brought about
by pairing correlations when going from an even to an odd nucleus.

A similar two step picture to go from even to odd systems has been defined by 
Balian, Flocard and Veneroni \cite{bal,flo} for the density operator. 
They introduced it in terms of number-parity-projected BCS states in the more 
general context of Fermionic super-conducting systems at finite temperature. 
However, they have not extensively studied the implications of this prescription 
in the context of nuclear structure. This intermediate vacuum has also been used 
as a natural definition of the smooth part of the microscopic binding energy in a
 work dealing with the OES~\cite{bend}.

The present work is organized as follow. In section~\ref{subsecbloc}, 
the standard perturbative qp creation process in BCS theory is reviewed and 
analyzed 
in order to point out some important characteristics  for the description 
of odd states. In section~\ref{subsecnew}, we propose a slightly different 
prescription for a perturbative treatment of odd nuclei. 
In section~\ref{subsechf}, the zero pairing limit of our revised picture is 
performed in order to show how it matches with the usual HF one.
In  section 3 and 4, detailed  HFB calculations are performed on even
and odd Ce isotopes in order to illustrate the procedures discussed at the 
perturbative level in section 2.
Conclusions are drawn in section~\ref{secconclu}.

\section{Odd nuclei description in a mean-field theory including pairing}
\label{secsimple}

\subsection{Perturbative nucleon addition process}
\label{subsecbloc}

Let us start with the BCS description of an even nucleus. For
a given effective Hamiltonian $\hat H$, one determines
the ground-state wave-function ${\rm \mid \Psi (N) > \ }$ with 
the constraint that it has a mean number of particles equal to $N$.
This constraint is imposed by the chemical potential $\lambda_N$
as Lagrange multiplier. 
A first approximation for the ground state of the odd neighbor 
with one more neutron
\footnote{In what follows we limit ourselves to the case of an odd isotope
with one more neutron. All what is presented can be easily
transposed to the removal of a neutron, or to
odd isotones.}
 is obtained by a perturbative one qp
creation
${\rm \mid \Psi_{k} > \ } = \ \alpha^{\dagger}_{k} \ {\rm \mid \Psi (N) > \ }$,
where $\alpha^{\dagger}_{k}$ is a qp
creation operator.
The average particle number of the state ${\rm \mid \Psi_{k} > \ }$ 
is $N+u^{2}_{k} - v^{2}_{k}$, where $v^{2}_{k}$ is the BCS occupation number of 
the state $k$. 
This average-number of particles
is not necessarily equal to $N+1$ and depends on the qp 
which has been selected~\cite{ring}. The energy difference between the 
state ${\rm \mid \Psi_{k} > \ }$ and the even ground-state is: 

\begin{eqnarray}
{\rm < \Psi_{k} \mid \ } \! \hat{H}  {\rm \mid \Psi_{k} > }\!-\!{\rm < \Psi (N) \mid \ } \! \hat{H} 
{\rm \mid \Psi (N) > } \! &=& \!  (u^{2}_{k}\!-\!v^{2}_{k}) \lambda^{N}\!+\! E^{N}_k \label{energydiff} \\
                    \!  &=& \! \frac{e_{k}(e_k\!-\!\lambda^{N})\!+\!{\Delta_k}^2}{\sqrt{ {(e_k\!-\!\lambda^{N})}^2\!+\!{\Delta_k}^2 }}  \nb \, \, ,
\end{eqnarray}
where the chemical potential $\lambda^{N}$ 
and the qp energy $E^{N}_{k}=\sqrt{ {(e_k - \lambda^{N})}^2 +{\Delta_k}^2 }$
are taken from the even ground-state. 

If the qp corresponds to a state having an energy
$e_k$ close to $\lambda^{N}$ ($u^{2}_{k} - v^{2}_{k} \approx 0$), 
the energy difference~\ref{energydiff} is approximatively equal to $E^{N}_k$ 
and is close to $\Delta_k$. 
However, the mean particle number is close to N, 
and ${\rm \mid \Psi_{k} > \ }$
is not a good candidate to describe an 
odd nucleus. To ensure an odd average number 
of particles in ${\rm \mid \Psi_{k} > \ }$, one should create a 
qp such that $(e_k - \lambda^{N})$ is much larger than $\Delta_k$. 
In such a case, the energy difference is approximatively given 
by $e_k \, \gg \lambda^{N}$. Once again, ${\rm \mid \Psi_{k} > \ }$ is 
not a good candidate for the ground state of the neighboring odd nucleus. 

This analysis shows that an odd nucleus wave-function cannot be approximated 
by a perturbative one qp excitation on the ground state of an even nucleus. 
Such a treatment does lead either to a wrong particle number and/or to
a bad energy. 

To circumvent the problem, one can put artificially
$u^{2}_{k} - v^{2}_{k} \approx \pm 1$ in Eq.~\ref{energydiff} 
(see for instance ref.~\cite{ring}, Chap. 6.3.4) which leads to:

\begin{equation}
{\rm < \Psi_{k} \mid \ } \! \hat{H}  {\rm \mid \Psi_{k} > }  -  {\rm < \Psi (N) \mid \ } \! \hat{H} 
{\rm \mid \Psi (N) > }  =  \pm \,  \lambda^{N}\!+\!\ E^{N}_k   \, \, .
\label{energydiff''}
\end{equation}

Such a	procedure is satisfactory for the determination
of energies. However, it does not provide  a tool
to calculate other observables as radii or deformation
since the wave-function still does not have the right mean particle number. 

Such a problem does not appear in self-consistent 
calculations since the one qp excitation is numerically performed together 
with a constraint on the correct average number of nucleons
\footnote{In BCS theory, it consists of solving the gap equation for an even 
number of particles excluding the state occupied by the odd nucleon.}. 
The chemical potential is readjusted  self-consistently whatever 
the starting point of the calculation is. 
However, to correct the inconsistencies of the perturbative 
picture is a necessary step to identify the various contributions to
the transition from an even nucleus to its odd neighbor.

\subsection{Revised perturbative scheme}
\label{subsecnew}

To improve the perturbative analysis  requires the definition
of a new vacuum on which qp states are blocked in such a way that
the choice of an energetically favorable qp leads to a state
with a nearly correct particle number.

A way to do that is to first determine the fully paired state having 
the right mean odd particle number. Let us denote that state by
${\rm \mid \Psi^{BCSE} (N\!+\!1)} >$. The subscript BCSE means 
that the state is constructed as an {\it Even} vacuum without qp 
excitation and without breaking time-reversal invariance 
but with an odd average particle
number. The lowest excitation energy with respect to this new reference vacuum
will  be generated by the qp with
the lowest energy but the ${\rm \mid \Psi^{BCS}_k >}$ will now have
an odd mean number of particles with a good approximation.
The energy difference $\Delta E(k)$ with the even neighbor becomes:

\begin{eqnarray}
&\begin{array}{ccccc}
\Delta E(k) &=& \underbrace{E^{BCSE}(N\!+\!1)-E^{BCS}(N)} &+&  E^{N\!+\!1}_{k}  \, \, \nb  \\
&& && \label{enerimpair-pair} \\
&\approx& \frac{\partial E^{BCSE}}{\partial N} &+&  E^{N\!+\!1}_{k}  \, \,  \, \, , \nb \\
\end{array}&
\end{eqnarray}

\noindent where $E^{N\!+\!1}_{k}$ is the energy of the lowest qp 
in ${\rm \mid \Psi^{BCSE} (N\!+\!1) > }$ and $E^{BCSE}(N)$ is the energy of the BCS fully paired vacuum with $N$ particle (even or odd). This result is formally 
similar to that of Eq.~\ref{energydiff''}. However, the 
qp excitation is now defined in the reference state 
${\rm \mid \Psi^{BCSE} (N\!+\!1)} >$ and no had hoc modification of the
chemical potential is required.
 
This procedure, although not perfect as it remains perturbative, is now 
qualitatively satisfactory from all points of view
and provides at the same time  a good approximation for the energy
and for the wave-function of an odd nucleus.

Such a  perturbative qp creation on top of the odd fully paired 
state, instead of the even neighbor's one, has already been introduced by 
Ring {\it et. al}~\cite{ring1} and used with success in Ref.~\cite{satula}. 
Its main justification was simplicity with respect to self-consistent blocking, 
but not the formal step achieved with respect to a perturbative 
qp creation performed on the even vacuum. 

\begin{figure}
\begin{center}
\leavevmode
\centerline{\psfig{figure=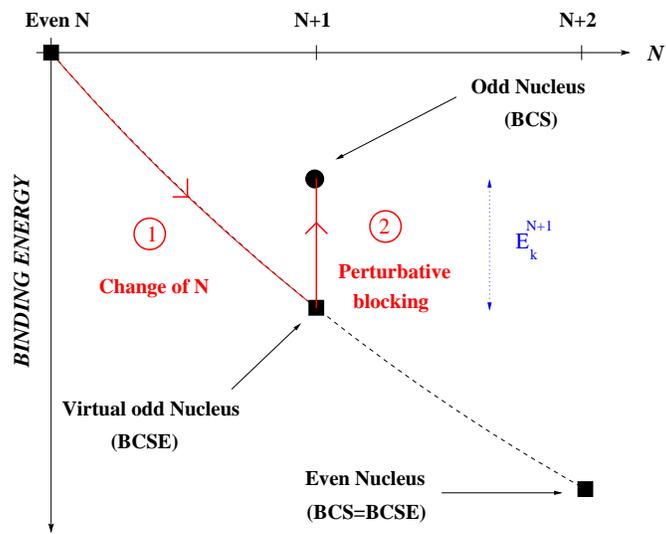,height=7cm}}
\end{center}
\caption{Schematic picture of the two step procedure
proposed to determine the ground state of an odd isotope.}
\label{revpicture}
\end{figure}

The introduction of an intermediate reference vacuum requires to
study  an odd nucleus in two steps.
This procedure, illustrated on Fig.~\ref{revpicture}, eliminates 
the inconsistency between the addition of a nucleon and the creation
of an energetically favorable qp excitation. From a mathematical point of view, 
it shows that the odd fully paired state is better grounded than an even neighbor 
ground-state as the zero-order reference for a perturbation theory of odd nuclei. 
In the rest of this paper, we will analyse these steps
from a physical point of view and use them to separate self-consistent calculations
in  two identified processes.

\subsection{Limit of zero-pairing}
\label{subsechf}

The description of an odd nucleus with respect to an even neighbor is at first 
sight 
less complicated in the absence of pairing. Indeed, there is no problem 
related to the particle number and an odd nucleus is simply obtained by
 adding a nucleon on the first empty level in the even neighbor's HF state. 
Two different approximations are used within this picture.

If time-reversal invariance is not broken, each single-particle
state is at least doubly degenerate and the odd nucleon is added using the 
filling approximation:
the first pair of empty levels in the even neighbor are identically occupied 
with probability 0.5 in the odd state\footnote{For spherical nuclei, one adds $1/2j+1$ particle in each state of the last degenerate j shell}.

If time-reversal symmetry breaking is properly
taken into account and for a deformed configuration, all
degeneracies are lifted and the first pair of empty
levels in the even isotope are occupied with probability 1 and 0
in the odd neighbor\footnote{For spherical nuclei, one orbital of the shell is completely filled, lifting the degeneracies. Several tries have to be done in order to get the lowest in energy.}.

Let us now analyze how the standard HF picture matches with the zero-pairing 
limit of the perturbative method described in section~\ref{subsecnew}. 
Most of the developments presented in this section have straightforward 
zero-pairing limits. Let us look explicitly to the limit for odd states only.

The limit of the perturbative one qp BCS state with an odd particle number is:

\begin{eqnarray}
&\begin{array}{lcccc}
{\rm \mid \Psi^{BCS}_{n} (N\!+\!1) > } &\rightarrow& {\rm \mid \Psi^{HF}_{n} (N\!+\!1) > } &=& a^{\dagger}_{n} \prod_{k=1}^{N/2} a^{\dagger}_{k} \, a^{\dagger}_{\bar{k}} \ket{0} \, \, , \label{limitodd1} \\
\end{array}&
\end{eqnarray}

\noindent whereas the fully paired odd vacuum leads to:

\begin{eqnarray}
&\begin{array}{lcccc}
{\rm \mid \Psi^{BCSE} (N\!+\!1) > } &\rightarrow& {\rm \mid \Psi^{HFE} (N\!+\!1) > } &=& \frac{1}{\sqrt{2}} \ \left( 1 +  a^{\dagger}_{n}\, a^{\dagger}_{\bar{n}} \right) \prod_{k=1}^{N/2} a^{\dagger}_{k} \, a^{\dagger}_{\bar{k}} \ket{0} \, \, . \label{limitodd2} \\
\end{array}&
\end{eqnarray}

One can check that:

\begin{equation}
 {\rm \mid \Psi^{HF}_{n} (N\!+\!1) > } = \alpha^{\dagger}_{n} \ {\rm \mid \Psi^{HFE} (N\!+\!1) > }
\label{excit}
\end{equation}
where $\alpha^{\dagger}_{n}= \frac{1}{\sqrt{2}} (a^{\dagger}_{n} - a_{\bar{n}})$ is the singular\footnote{Other qp operators $\alpha^{(\dagger)}_{k}$ ($k \neq n, \bar{n}$) tend to standard particle creation or annihilation operators $a^{(\dagger)}_{k}$.} zero-pairing limit for the lowest qp creation operator. 

The wave function ${\rm \mid \Psi^{HFE} (N\!+\!1)> }$ introduced 
as the limit of the BCSE state is none of the two currently used HF
wave-functions. However it leads to the same 
one-body density matrix, and thus to the same energy as the HF wave-function~\footnote{The filling
approximation is actually defined through a density operator which is a statistical 
mixture of the two Slater determinants where one of the two time-reversed 
orbitals at the Fermi energy is filled. The ${\rm \mid \Psi^{HFE}> }$ state~\ref{limitodd2} for odd nuclei is a linear combination of the two neighboring even HF states.} obtained using the filling approximation.

The HF ground-state for odd nuclei is now described by
a {\it one qp excitation} on top of the HFE state
and not as in the usual procedure
directly on top of the HF wave function of an even neighbor through particle operators.
The two-step picture defined in the BCS case is thus  extended 
to the zero pairing limit
and will allow an analysis of the OES for 
any pairing correlations intensity.

The zero pairing limit illustrates the physical content of the nucleon
addition process. The nucleon is {\it added} in the HFE wave function 
by increasing the occupation of each state of the last couple of degenerate 
orbits by 0.5.
For odd $N$, the qp excitation {\it specifies} which one of the two states 
will eventually be occupied by the single nucleon in the odd wave function. 
The only difference in presence of pairing is that the nucleon is added over 
the whole fermi sea in the BCSE wave function because of pair scattering, while 
the qp creation still specifies the state eventually occupied by the single 
nucleon.

\subsection{Self-consistent HFB treatment of odd nuclei}
\label{secself}

Since time-reversal symmetry is broken in an odd nucleus, a
proper treatment of pairing correlations requires the use of the HFB method
and the introduction of time-odd components in the mean-field.

In this context, Eq.~\ref{enerimpair-pair} is replaced by:

\begin{eqnarray}
E^{HFB}(N) &=& E^{HFBE}(N) + [E^{HFB}(N) - E^{HFBE}(N)] \nb  \\
&=& E^{HFBE}(N) +  \, \, \, \,  \overbrace{E^{pol}(N) \, \, \, \, \, + \, \, \, \, \, \, \, \Delta(N)\,}  \label{defener} \, \, \, \, \, \, ,
\end{eqnarray}
where HFBE refers to fully paired states, $\Delta(N)$ is the 
positive contribution due to the self-consistent blocking of 
pairing correlations in odd nuclei due to the presence of a single nucleon. $E^{pol}(N)$ is the part of the binding energy related to polarisation effects in odd nuclei. First, it contains a static 
deformation-polarisation of the core induced by the non-zero multipole moment 
of the odd nucleon density. Second, the breaking of the time-reversal symmetry by this 
odd nucleon brings about non-zero spin and current contributions. 
The sum $E^{pol}(N) + \Delta(N)$ can be viewed as the self-consistent 
qp energy to be compared to the perturbative one, $E^{N\!+\!1}_{k}$. 

In a fully self-consistent calculation, HFE and HF states are 
defined as the self-consistent zero-pairing limit of HFBE and HFB states. 

\section{Results}
\label{secres}

\subsection{Addition of a nucleon}
\label{subsecadd}

In this section, we  apply the decomposition of energy introduced
in section \ref{secsimple} to a chain of cerium isotopes, 
from $^{118}$Ce to $^{166}$Ce. 
Our aim is to determine to which extent this decomposition allows
to decouple both effects related to the
addition of a nucleon. 

We have performed Hartree-Fock-Bogolyubov 
plus Lipkin-Nogami (HFBLN) calculations with the
formalism and forces in the particle-hole 
(SLy4 Skyrme force) and particle-particle 
(zero-range density dependent pairing force) channels 
described in ref \cite{tera,rigol}. Each odd nucleus is 
calculated twice: first, 
as a HFBLN fully paired vacuum with an odd average number of neutrons 
(HFBE state) and then with the fully self-consistent 
HFBLN scheme (HFB state). In this case, several qp configurations are investigated  
to determine the one corresponding to the ground state.

\begin{figure}
\begin{center}
\leavevmode
\centerline{\psfig{figure=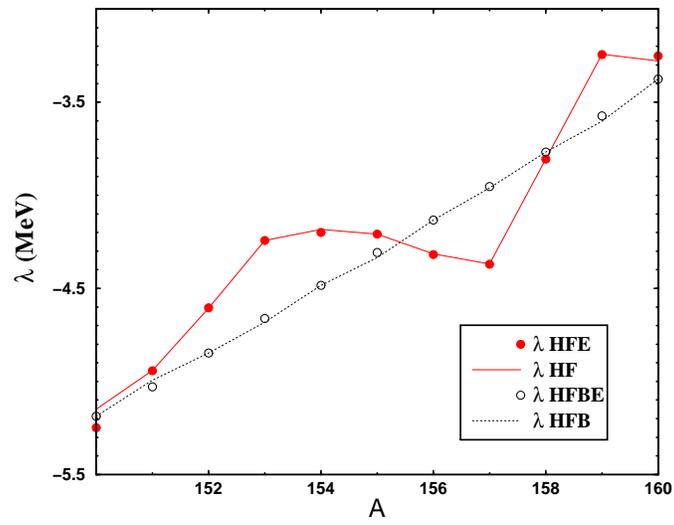,height=7cm}}
\end{center}
\caption{Chemical potential along the cerium isotopic line between $^{150}$Ce and $^{160}$Ce for HFE, HF, HFBE and HFB calculations.}
\label{fermi}
\end{figure}

Cerium isotopes have been chosen because of their intermediate masses 
and also because their axial mean quadrupole deformation evolves regularly 
along the whole chain from 
sphericity to large deformations.

Fig.~\ref{fermi} displays the chemical potential as a function of mass 
from $^{150}$Ce to $^{160}$Ce for HFE, HF, HFBE and HFB calculations. 
The chemical potentials in 
HF(E)~\footnote{The parenthesis in HFB(E) means that the corresponding 
sentence deals with ``HFB {\it and} HFBE''. Identically, 
HF(B)E means ``HFE and HFBE'' and HF(E) means ``HF and HFE''.}
 states are well defined in the zero-pairing limit of HFB(E) ones. 

The results are represented for a sub-zone which is representative of the full cerium isotopic line that we have calculated.
The figure shows that the
change of Fermi level due to the addition of a nucleon is fully
taken into account by introducing only a constraint on an odd
particle number (HF(B)E calculations)
and is not affected by the self-consistent blocking in the final state. 
It proves that the qp creation carries no additional particle with respect to the reference vacuum HF(B)E, as expected from the perturbative picture. 
This justifies from a quantitative point of view the decoupling of the 
single nucleon addition in the fully paired vacuum and its blocking in the full HF(B) odd state.

\begin{figure}
\begin{center}
\leavevmode
\centerline{\psfig{figure=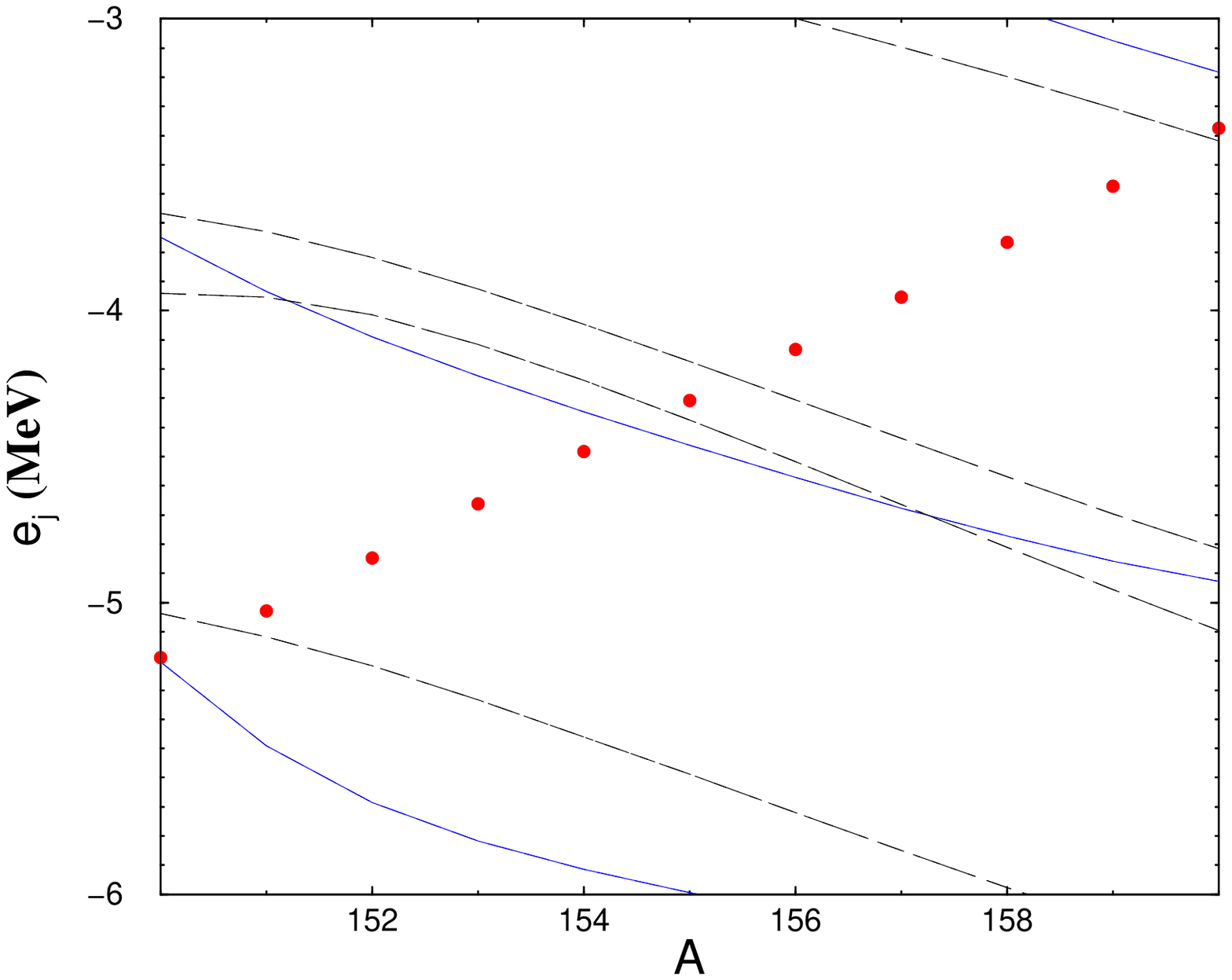,height=6.83cm} 
\psfig{figure=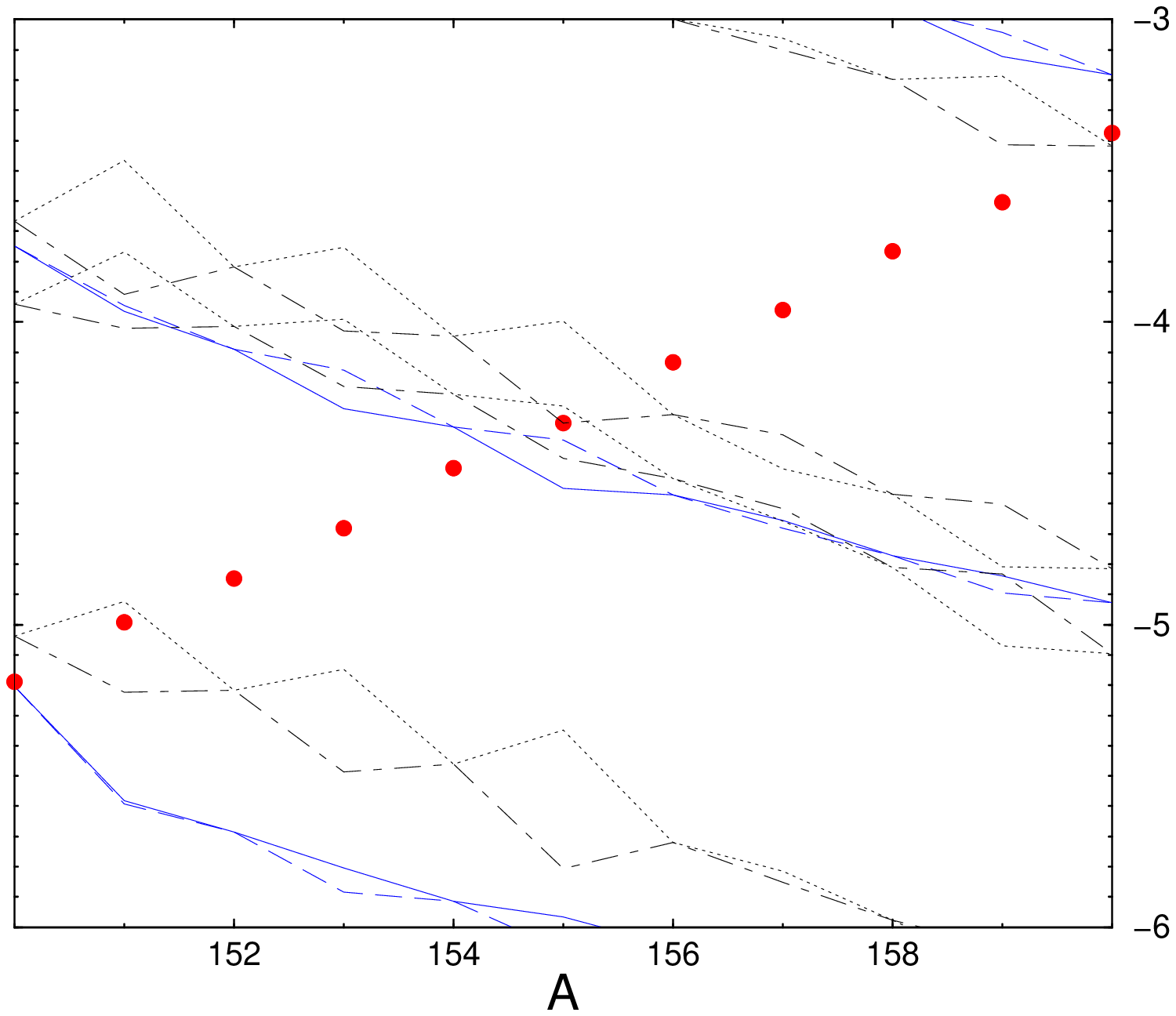,height=7cm}}
\end{center}
\caption{Single-particle spectrum as a function of the mass number $A$ 
between $^{150}$Ce and $^{160}$Ce. The left panel corresponds to the HFBE case, 
the right one to the HFB one.
 The conventions for (parity, signature) are~: 
(+,+)=solid line, (+,-)=long-dashed line, (-,+)=dotted line, (-,-)=dot-dashed line. Circles are for the chemical potential.}
\label{spectrehfb}
\end{figure}

\begin{figure}
\begin{center}
\leavevmode
\centerline{\psfig{figure=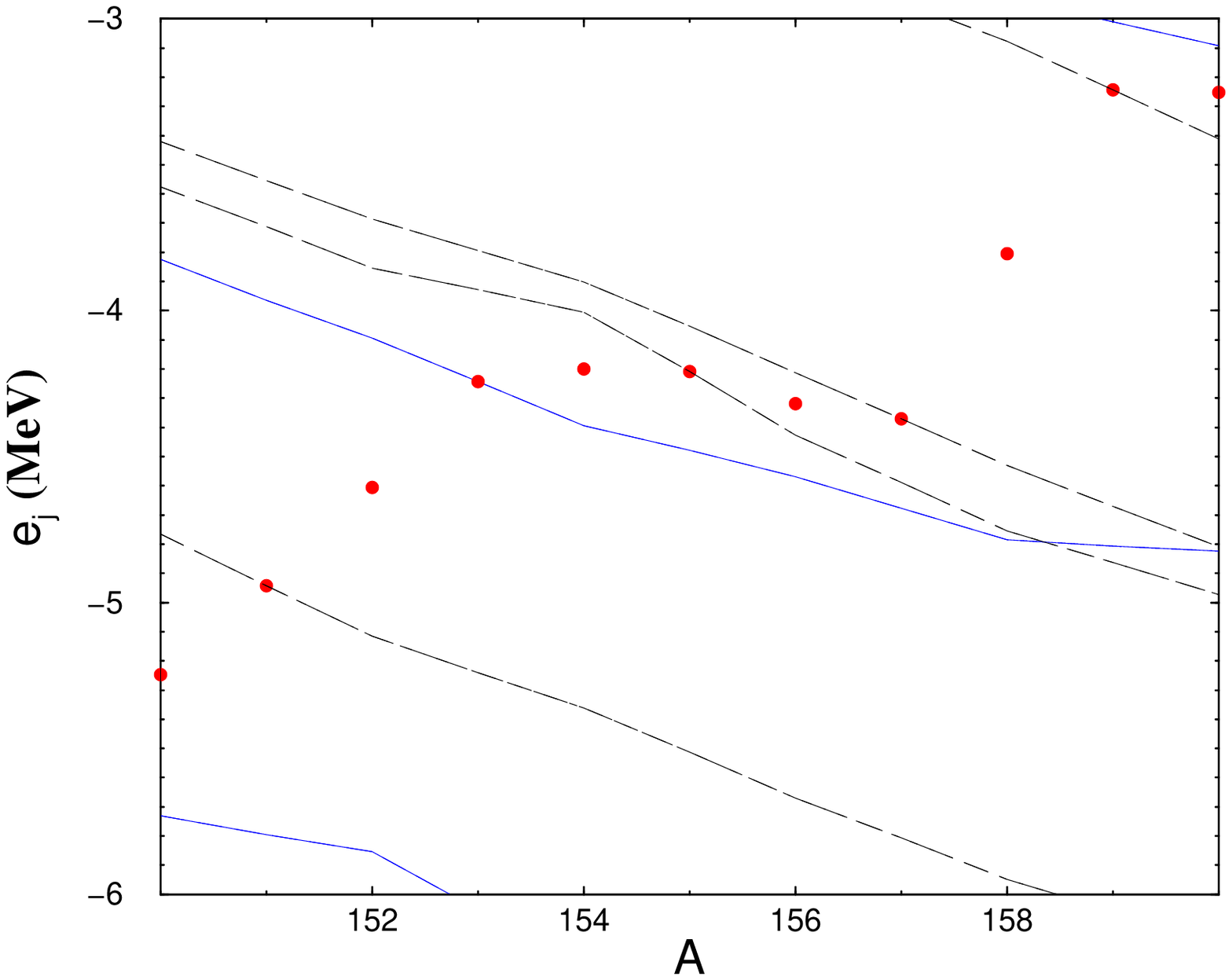,height=6.83cm} \psfig{figure=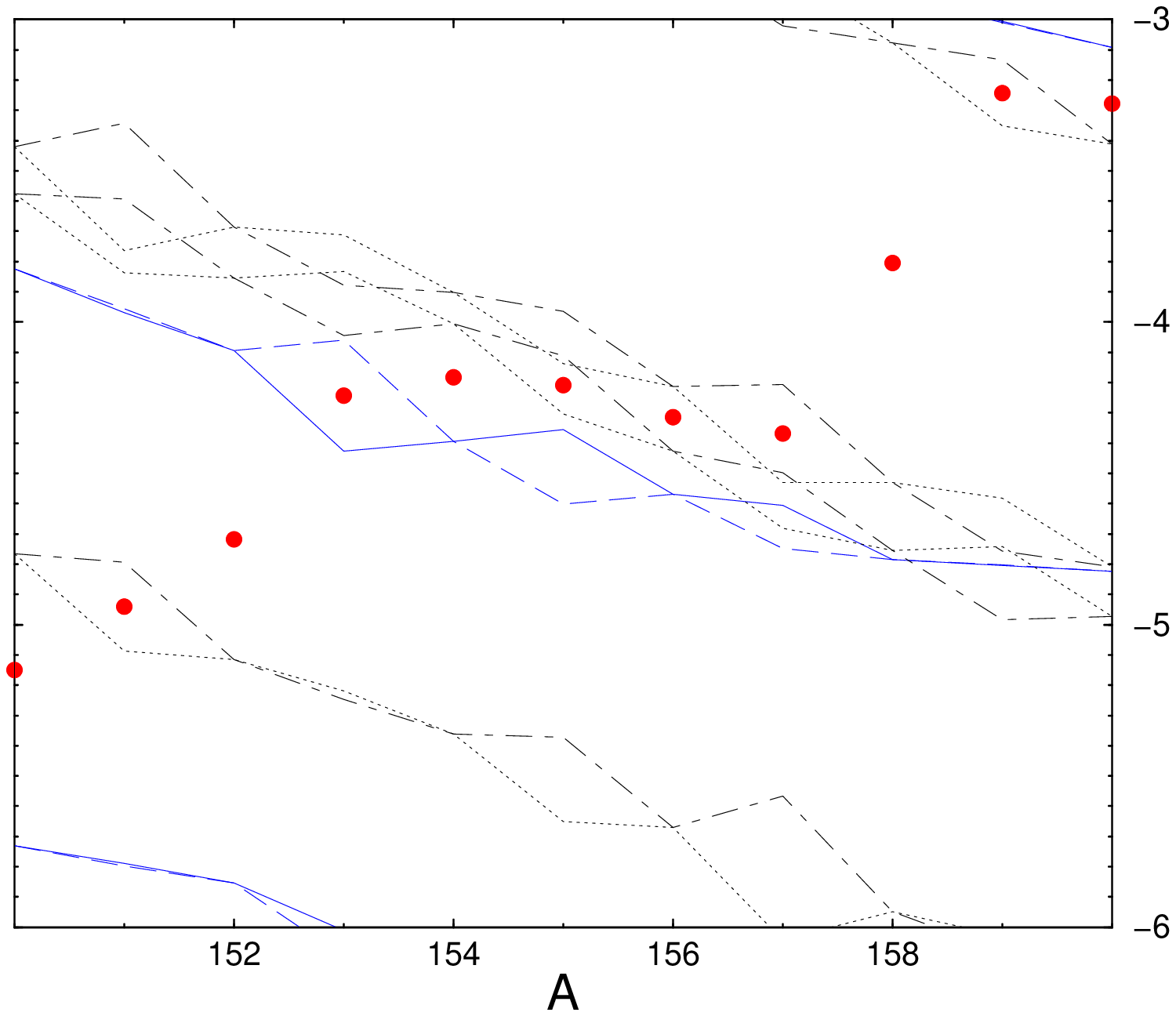,height=7cm}}
\end{center}
\caption{Same as Fig. \ref{spectrehfb} for the HFE and HF cases respectively.}
\label{spectrehf}
\end{figure}

The evolution of $\lambda$ with N depends on the underlying mean-field 
as well as on the occupation numbers. On Fig.~\ref{spectrehfb} 
are shown the neutron single-particle spectra obtained in the HFBE and HFB 
calculations. For odd nuclei, the double degeneracy of the single-particle
energies is lifted in the HFB calculation, leading to an odd-even
staggering of these single-particle energies. However, if one takes
the mean energy between the states of a doublet, the HFBE and HFB 
single-particle spectra are identical. 
Fig. \ref{spectrehf} displays the single-particle 
neutron spectra for HF(E) calculations and shows that
the same  remark remains valid in the zero-pairing limit. 

We can conclude from these comparisons that 
constraining the HF(B)E state to an odd number of particles
without creating a qp excitation lead to the same mean-field 
as the full HF(B) state, except for small polarisation effects
due to the breaking of time-reversal invariance.

We will therefore take  E$^{HF(B)E}$ as 
the ``Mean-Field'' 
part of the binding energy. In the zero-pairing limit, this definition 
reduces to the time-reversal invariant part of the interaction.
When pairing correlations are present, 
this energy includes also the part of the pairing energy
which is not related to the blocking effect and which varies smoothly 
with the particle number.

\begin{figure}
\begin{center}
\leavevmode
\centerline{\psfig{figure=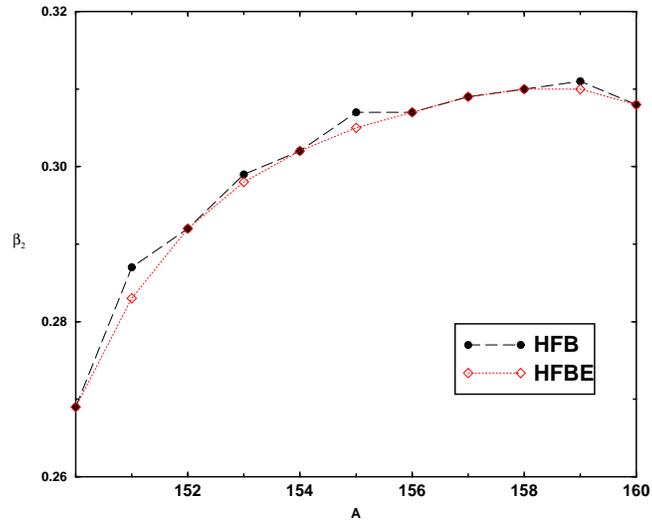,height=7cm}}
\end{center}
\caption{HFB and HFBE axial quadrupole deformation parameters 
$\beta_2 = \sqrt{\frac{\pi}{5}} \frac{< \hat{Q}_{20} >}{A < r^2 >}$ 
along the cerium isotopic line between $^{150}$Ce and $^{160}$Ce.}
\label{beta2}
\end{figure}

Even if the previous conclusions are valid in both the HF(E) and HFB(E) cases, the situation differs depending whether pairing correlations are included or not. The left panels of Fig.~\ref{spectrehfb} and~\ref{spectrehf} show that the single-particle spectra are different in the two cases. In addition, Fig.~\ref{fermi} shows the smoother behavior for $\lambda_{HFB(E)}$ as compared to $\lambda_{HF(E)}$. This proves that the inclusion of the pairing is not a perturbative effect and deeply modifies the wave-function. 

The fundamental difference between ${\rm \mid \Psi^{HFBE} > }$ and ${\rm \mid \Psi^{HFE} > }$ is related to the way a nucleon is added in both cases: while it is added on a specific pair of time reversed orbits in the HFE case as discussed in section~\ref{subsechf}, it is spread out on several
orbits around the Fermi level in the HFBE case because of pair scattering, making the variation of $\lambda_{HFB(E)}$ smoother with A.

The same type of analysis is valid for other observables. Fig. \ref{beta2} 
gives  the mass number dependence of the axial quadrupole moment.
The smooth variation related to the modification of the mean-field is fully taken into account in the HFBE state. In the HFB calculation of odd nuclei, a tiny additional change of deformation appears due to the qp creation. The proposed scheme allows to decouple the two contributions.

\subsection{Qp creation effect without pairing}
\label{subsecblockingHF}

We have studied the process of the addition of one nucleon through the 
definition of ${\rm \mid \Psi^{HF(B)E} > }$ as a reference vacuum. 
Let us focus now on the blocking of this added nucleon. We propose a simple tool to 
disentangle the two components of the qp creation process: 
the breaking of time-reversal invariance and the quenching of pairing. 
First, the zero-pairing case is treated because it contains one of 
the two effects only. 

\begin{figure}
\begin{center}
\leavevmode
\centerline{\psfig{figure=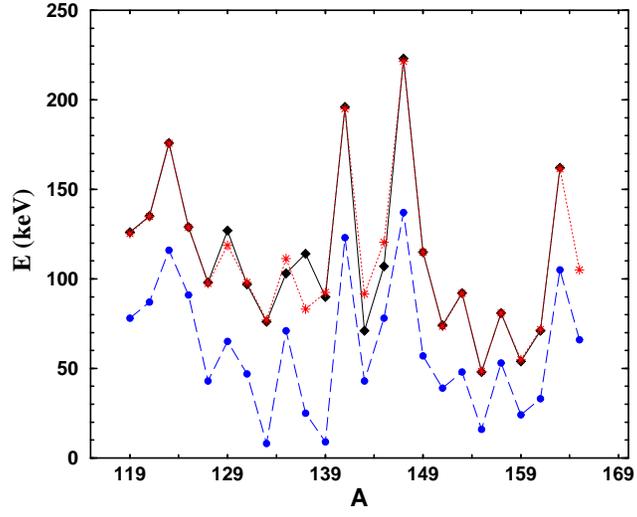,height=7cm}}
\end{center}
\caption{Diamonds: energy difference E$^{HF}$ - E$^{HFE}$ (see text) along 
the cerium isotopic line. Stars: approximation 
$[\tilde{e}_{n}^{HF}-\tilde{e}_{\bar{n}}^{HF}]/4$ for the polarisation 
effect in odd nuclei. 
Circles: time-odd mean-field terms contribution 
$ -\bar{V}^{odd}_{n\bar{n}n\bar{n}}/4 \,$ to the time-reversal symmetry breaking effect.}
\label{pol}
\end{figure}

The energy difference E$^{HF}$ - E$^{HFE}$ displayed on Fig.~\ref{pol}
gives a direct  information on polarisation effects brought by
the  odd nucleon, especially through the breaking of the time-reversal symmetry.
 This symmetry breaking removes the degeneracy between 
signature partners (see Fig.~\ref{spectrehfb} and \ref{spectrehf}). 
As noticed in previous works~\cite{PH1,PH2}, this effect is the largest 
for signature partners corresponding to the qp which is created. 
Fig.~\ref{pol} shows that the net effect is repulsive and of the order 
of a few hundreds keV. Along the cerium isotopic line,
it ranges from 48 keV for $^{155}$Ce to 223 keV 
for $^{147}$Ce. 

The polarisation effects obtained with the Sly4 interaction 
have been found attractive on average in ref~\cite{satupol}
for light nuclei. 
The difference with our results may be related either 
to a mass dependence of the effect or to a 
competition between isovector and isoscalar effects~\cite{satupol2}. 
The effect of the isovector terms of the interaction is indeed very
weak in the study of N $\approx$ Z nuclei of Ref.~\cite{satupol} 
while it is not the case in the present study of Ce isotopes.

In appendix A is derived an approximate expression for the difference between
HF and HFE energies. It is based on the assumption that the HF and HFE 
single-particle wave-functions are identical, leading to the same
matrix elements for the two-body force; the two N-body wave-functions only differing through individual occupation numbers. This assumes that the deformation-polarisation of the core induced by the blocked nucleon is very weak. 

This perturbative calculation for the polarisation effect in absence of pairing gives:

\vspace{0.1cm}

\begin{equation}
E^{pol} = E^{HF}-E^{HFE} \approx \frac{\tilde{e}_{n}-\tilde{e}_{\bar{n}}}
{4} \, \, \, \, \, \, \, \, \, \, \, .
\label{pol2body}
\end{equation}
\vspace{0.1cm}

\noindent where  $\tilde{e}_{n}$ and $\tilde{e}_{\bar{n}}$ are the  split 
orbits in the HF wave-function having occupation numbers 
1 and 0 respectively.

The results obtained using this approximation are also plotted  on Fig.~\ref{pol} and are in very
good agreement with the full polarisation effect along the 
whole cerium isotopic line. 
This justifies that the individual wave-functions are  marginally modified by 
the qp creation and shows that core deformation-polarisation effects are weak.

The energy difference \ref{pol2body} can be rewritten in terms of a single 
unperturbed diagonal particle-hole matrix element of the two-body 
force between the blocked time-reversed states: 

\begin{equation}
E^{pol} \approx - \frac{\bar{V}^{p-h}_{n\bar{n}n\bar{n}}}{4} 
\, \, \, \, \, \,  \, \, \,  \, \, .
\label{matelem}
\end{equation}
\vspace{0.1cm}

On Fig.~\ref{pol} is also plotted the contribution of the time-odd terms of the interaction $-\bar{V}^{odd}_{n\bar{n}n\bar{n}}/4 \,$ to E$^{pol}$. The explicit expression of $\bar{V}^{odd}_{n\bar{n}n\bar{n}}$ 
can be worked out from Ref.~\cite{engel}. 
One can see that the time-odd terms
are roughly responsible for 2/3 of the time-reversal symmetry breaking effect
for all isotopes. The
erratic behavior of the polarisation effect as a function $A$ is directly related 
to these terms, while the time-even terms seem to be less sensitive 
to the characteristics of the blocked orbits.

Eq. \ref{pol2body} and \ref{matelem} allow to extract the polarisation 
effect in a simple way from a single calculation. Either one performs 
a full HF calculation and evaluates the polarisation effect in term 
of the single-particle energy splitting, or one performs a simpler HFE 
calculation and evaluates the polarisation effect by extracting the 
relevant matrix element. 

The single-particle character of the polarisation energy illustrated by Eq. \ref{pol2body} and \ref{matelem} has been pointed out in ref. \cite{satupol} where it has been shown 
that:

\begin{equation} 
E^{pol}(N-Z=2n) = E^{pol}(N-Z=2n+1) + E^{pol}(N-Z=2n-1) \, \, \, \, \, \, .
\label{polsingpart}
\end{equation}
 
The fact that the polarisation energy is shown to be related to the splitting of 
a single pair of states or a single matrix element simplifies the analysis. 

One can therefore relate the magnitude of the polarisation energy in an odd nucleus with three properties of the blocked orbital.
In decreasing order of importance they are:  
a small $j_{z}$ component on the deformation axis (K quantum number), 
a down-slopping behavior of the individual energy with $A$ and 
a large total angular momentum $j$ for the spherical shell
from which the orbit originates. 
That orbitals with these characteristics have large polarisation effects
is not surprising since the same orbitals are known to be 
very sensitive to rotation which is also an effect related to time-odd terms 
of the mean-field. The large energy difference that can be seen
on Fig.~\ref{pol} for $^{147}$Ce is 
associated with the blocking of the very down-slopping Nilsson 
orbital  [660]1/2, originating from the 1$i$13/2 shell. In $^{123}$Ce and 
in $^{141}$Ce, the large polarisation corresponds to the  [541]1/2 blocked 
orbitals coming from the 2$f$7/2 spherical shell. These last two nuclei 
are of particular interest because although they have  different masses 
and very different deformations (see Fig.~\ref{beta2}), their large 
polarisation energy is of the same order of magnitude 
since it is related to the matrix element involving the same pair of Nilsson 
blocked states. This demonstrates that K is a relevant quantum number in 
order to characterize the magnitude of E$^{pol}$. These 
conclusions are valid along the whole cerium isotopic line.

\subsection{Qp creation effect with pairing}
\label{subsecblockingHFB}

When pairing correlations are included, 
the energy difference E$^{HFB}$ - E$^{HFBE}$ mixes both 
the effect of the blocking of pairing and the polarisation effect and cannot be used to extract one of them only. However, it is shown in Appendix A that the approximation~\ref{pol2body} 
for the polarisation effect still holds for HFB calculations. 

\begin{figure}
\begin{center}
\leavevmode
\centerline{\psfig{figure=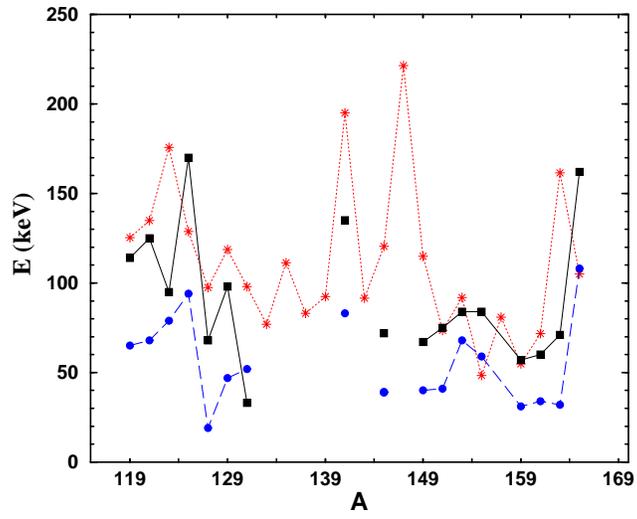,height=7cm}}
\end{center}
\caption{Stars: approximation $[\tilde{e}_{n}^{HF}-\tilde{e}_{\bar{n}}^{HF}]/4$ 
for polarisation effect in odd nuclei. Squares: same as stars for the
HFB calculation.  Circles: time-odd mean-field terms 
contribution $ -\bar{V}^{odd}_{n\bar{n}n\bar{n}}/4 \,$ to the polarisation effect.}
\label{polhfb}
\end{figure}

As a consequence, E$^{pol}$ has the same order of magnitude in average 
in HF and HFB cases although it can be significantly different for a given nucleus
as it can be seen on Fig.~\ref{polhfb}. Polarisation energies in the non zero pairing case are given on Fig. 7 only for nuclei for which the assumption of a perturbative calculation is valid.

Due to self consistency, the ordering of the qp levels may
be different in the HFE and HFBE calculations.
 On Table 1 are listed the Nilsson labels of the orbitals 
corresponding to the qp created for each odd cerium isotopes in
the HF and HFB calculations. 
The differences that can be seen on Fig.~\ref{polhfb} 
between both calculations
are  related to the different qp corresponding to the ground states of
 $^{123,145,147,155,163,165}$Ce. 
For $^{161,163,165}$Ce, the difference is due
to a shift by two neutron units for the blocked qp. Thus, the polarisation energies in $^{163}$Ce and$^{165}$Ce with pairing included are respectively equal to those of $^{161}$Ce and $^{163}$Ce
without pairing.

>From $^{141}$Ce to $^{163}$Ce the polarisation energy has
a smoother behavior when pairing is included.
In the HF calculation,  the lowest qp automatically changes
from one odd nucleus to next, while the same qp may remain the
lowest in energy in several neighboring nuclei in the HFB case, making the effect smoother with A.

There are only eight nuclei for which the created qp excitation is identical in 
both calculations. For five of them, $^{119,129,151,159}$Ce, 
the polarisation effect is not significantly modified by the inclusion 
of pairing. However, for three others, $^{125,131,141}$Ce,
the self-consistency between the mean-field and the pairing field in
the HFB calculation is large enough to induce a significant modification of
the energy splitting $\tilde{e}_{n}-\tilde{e}_{\bar{n}}$.

As in zero-pairing case, the contribution of time-odd terms, plotted on 
Fig.~\ref{polhfb},  contributes  to approximately 2/3 of
the full polarisation energy and follows the average behavior of $E^{pol}$. 
This illustrates the sensitivity of time-odd components of the 
interaction to the ($j_{z}, j$) quantum numbers of the created qp.

The polarisation energies have been calculated for the tin isotopic line 
in Ref.~\cite{rutz} within the framework of relativistic mean-field 
theory.  Pairing correlations were treated in the BCS approximation.
It was found  attractive 
by about 300 keV for two different parametrizations of the force. 
We have performed a similar study within our approach and 
we have found that the polarisation energy for the same isotopes is
repulsive by approximately 100 keV. These contradicting
results show the necessity of further analysis on the 
dependency of the polarisation effect on self-consistent models and forces
that are used. The main
source of uncertainty is due to the fact that the time-odd 
terms of the phenomenological forces 
have only been indirectly constrained through time-even ones~\cite{doba}. 
The possibility to associate the polarisation effect with a specific matrix element 
of the force could open a way for the inclusion of a constraint on these 
terms in the standard fitting procedures.

\section{Conclusions}
\label{secconclu}

In this paper, we have re-analyzed the way odd nuclei are described
in self-consistent mean-field calculations with a double goal.
We wanted to focus on the nucleon addition process in the nuclear mean-field wave-function and on its energetic consequences when going from one nucleus to the neighbor. We also wanted to
define a procedure such that the HF treatment of odd nuclei
is the zero-pairing limit of the HFB treatment. To achieve these
goals, we have defined a two steps procedure. 

The first step corresponds to the description of odd nuclei as even 
ones, by an appropriate constraint on the particle number.
This has required a  modification of the usual HF filling approximation. 
It has been shown that this pseudo even state takes into account the variation of the mean-field with mass number.

In the second step, a qp is created on top
of this odd vacuum. This does not modify the position of the Fermi level whatever the characteristics of the created qp (shell, spin, parity). 
Thus, the physical effect of the addition of 
a nucleon is contained in this fully paired state while the qp creation
brings the extra polarisation and the modification of pairing
correlations due to the non pairing of this added odd nucleon.

An application to the Ce isotopic line
has illustrated the relevance of this decoupling and the possibility
to remove a smooth behavior of physical observables like energies and
quadrupole moments thanks to this pseudo even structure.

Thus, the creation of a qp has two effects on binding energy of odd nuclei. The first
is related to the breaking of time-reversal invariance while
the second is due to the non pairing of a nucleon.
In the HF case, we have shown that the first effect
can be related with an excellent accuracy to the lift of the Kramers degeneracy of the 
conjugate pair of orbits in the single-particle spectrum having occupation 1 and 0. This effect
is also present with pairing correlations and can still be related to the same Kramers degeneracy. In this case, its energetic effect is dominated by the quenching of pairing in the qp creation process.
We have also shown that the first effect can be associated
with a single matrix element in the particle-hole channel. 
This result is promising for the necessary inclusion of a constraint 
on the mean-field time-odd terms in the standard fitting procedure of 
phenomenological two-body forces.
In the present study, this result has allowed us to isolate the specific 
contributions of these time-odd terms which have been shown to account 
for 2/3 of the time-reversal symmetry breaking effect in odd cerium isotopes 
when using SLy4.

In the second part of this work, we will use the above analysis of odd nuclei in order to understand the different contributions to the odd-even mass fomulas currently used to approximate the pairing gap~\cite{mad,mol}.

\vspace{1cm}

{\bf APPENDIX: Perturbative calculation of the qp creation effect}
\vspace{0.6cm}

The energy of an HFB state can be  expressed
in the canonical basis and is given by:

\begin{equation}
E^{HFB} = \sum_{k} \, \left( e_{k} - \frac{1}{2} \, \sum_{k'} \, \bar{V}^{p-h}_{kk'kk'} \, v^{2}_{k'} \right) v^{2}_{k} - \frac{1}{4} \, \sum_{k,l} \, \bar{V}^{p-p}_{k\bar{k}l\bar{l}} \, u_{k} \, v_{k} \, u_{l} \, v_{l}  \, \, \, \, \, \, \, \, ,
\label{HFB}
\end{equation}\vspace{0.3cm}

\noindent where $\bar{V}^{p-h}_{kk'kk'}$ and 
$\bar{V}^{p-p}_{k\bar{k}l\bar{l}}$ are  the antisymmetrised matrix 
elements of the two-body force, $v^{2}_{k}$ are the eigenvalues 
($u^{2}_{k} = 1 - v^{2}_{k}$) of the density matrix and $e_{k}$ the 
diagonal matrix elements of the Hartree-Fock field in the
canonical basis:

\begin{eqnarray}
e_{k} &=& t_{k} + \sum_{k'} \,\bar{V}^{p-h}_{kk'kk'} \, v^{2}_{k'} \, \, \, \, \, \, \, \, ,
\label{individual}
\end{eqnarray}

\noindent where $t_{k}$ is the diagonal matrix element of the kinetic energy 
operator for an individual wave-function $\phi_{k}$ in 
the canonical basis. 
We use the convention that $k$ and
$\bar{k}$ are paired partners. If not specified, the sum runs 
over all individual states. For simplicity, the rearrangement terms in the mean-field due to the
density dependence of the Skyrme interaction are not included. However,
their introduction does not modify the final expression for the polarisation
energy.

The diagonal matrix element of the pairing field can also be defined by:

\begin{equation}
\Delta_{k}  = - \frac{1}{2} \, \sum_{l} \, \bar{V}^{p-p}_{k\bar{k}l\bar{l}} \, u_{l} \, v_{l}
\label{gap}
\end{equation}\vspace{0.3cm}

Let us consider two approximations $a$ 
and $b$~\footnote{All quantities referring to the case $b$ will be upper-lined with a tilde.} 
of the exact HFB state of
a given nucleus. To evaluate the difference  between
the energies obtained with these two approximations, we will suppose
that one has only to take into account the changes in occupation 
$v^{2}_{k}$ and that changes in the canonical basis wave functions
can be neglected in the calculation of the matrix elements of the
interaction.

Let us take as state $a$ a time-reversal invariant HFB state. 
One has then $e_{k} = e_{\bar{k}}$.  The differences between the 
individual energy ${e}_k$ in the state $a$ and the energies $\tilde{e}_k$ and $\tilde{e}_{\bar{k}}$ in the state $b$ are given by:

\begin{eqnarray}
\tilde{e}_{k} - e_{k} &=& \sum_{k'} \, \bar{V}^{p-h}_{kk'kk'} \left( \tilde{v}^{2}_{k'} - v^{2}_{k'} \right) \, \, \, \, \, \, \, \, \, , \nb \\
\tilde{e}_{\bar{k}} - e_{k} &=& \sum_{k'} \, \bar{V}^{p-h}_{\bar{k}k'\bar{k}k'} \left( \tilde{v}^{2}_{k'} - v^{2}_{k'} \right) \, \, \, \, \, \, \, \, \, ,
\label{individualdiff2}
\end{eqnarray}
\vspace{0.3cm}

Using the relations $\bar{V}^{p-h}_{kk'kk'} = \bar{V}^{p-h}_{k'kk'k}$ 
and  $\sum_{k'} \, \bar{V}^{p-h}_{kk'kk'} \,  v^{2}_{k'} = \sum_{k'} \, 
\bar{V}^{p-h}_{k\bar{k'}k\bar{k'}} \, v^{2}_{k'}$, 
one can derive the following expression:

\begin{eqnarray}
E^{b} - E^{a} &=& \frac{1}{2} \, \sum_{k} \, \left[  \, \left(  e_{k} + \tilde{e}_{k} \right) \, \left( \tilde{v}^{2}_{k} - v^{2}_{k} \right) \, - \, \tilde{\Delta}_{k} \, \tilde{u}_{k} \, \tilde{v}_{k} \, + \, \Delta_{k} \, u_{k} \, v_{k} \right]  \, \, \, \, \, \, \, \, .
\label{diffener}
\end{eqnarray}

\begin{appendix} 

\section{Without pairing}
\label{withoutpairing}

Let us take for $a$ the HFE wave-function and for $b$ the HF one.
 The energy difference $E_b-E_a$ is due to time-reversal symmetry 
breaking and is equal to the polarisation energy E$^{Pol}$.
The occupation numbers of all individual states below the
Fermi level $\lambda^{odd}$ are 1, and 0 for all states above, 
except for the pair of states with energies just above $\lambda^{odd}$ which are $1/2$ for $a$
and 1 and 0 for $b$.

The energy difference given by Eq.~\ref{diffener} becomes:

\begin{eqnarray}
&\begin{array}{lcccl}
E^{Pol}  &=& E^{HF}-E^{HFE} &=& \left( \tilde{e}_{n}-\tilde{e}_{\bar{n}}  \right)  \, / \,4  \, \, \, \, \, \, \, \, \,  \, \, , \\
       &&        &=&  - \bar{V}^{p-h}_{n\bar{n}n\bar{n}} \, / \,4 \, \, \, \, \, \, \, \, \,  \, \, \, \,  \,  ,
\label{HF_HFE}
\end{array}&
\end{eqnarray}
\vspace{0.3cm}

\noindent the second expression being obtained using 
Eq.~\ref{individualdiff2} and the cancelation of the 
antisymmetrised matrix element $\bar{V}^{p-h}_{nnnn}$.

\section{With pairing}
\label{withpairing}

In the presence of pairing correlations, the energy difference
(E$^{HFB}$ - E$^{HFBE}$) contains contributions coming from the 
blocking of pairing and from polarisation effects due to the breaking
of time-reversal invariance. 
This energy difference is calculated in two steps. First,  
the polarisation effects are eliminated by performing a 
filling approximation. This means that starting from the fully paired HFBE state, the occupation probabilities in the canonical basis are changed from 
$v^{2}_{k}$ to  $\tilde{\tilde{v}}^{2}_{k}$ for all states, except for
two of them close to the Fermi energy for which the occupations  $\tilde{\tilde{v}}^{2}_{n}$
and $\tilde{\tilde{v}}^{2}_{\bar n}$ are set to 1/2.
The blocking of pairing is taken
into account by excluding these two states $n$ and $\bar{n}$
from the calculation of $\Delta$.
Second, we consider the fully blocked state, for which the occupation
probabilities are denoted by $\tilde{v}^{2}_{k}$, the state $n$ and $\bar{n}$
having an occupation 1 and 0 respectively.

We have checked numerically that it is a fair approximation 
to take the occupation probabilities $\tilde{v}^2_k$ and 
$\tilde{\tilde{v}}^{2}_{k}$ and the pairing gaps 
$\tilde{\tilde{\Delta}}_{k}$ and
$\tilde{\Delta}_{k}$
equal for all $k$ and $\bar{k}$ except for $n$ and
$\bar{n}$.

Using this assumption, one can show that:

\begin{eqnarray}
&\begin{array}{lcccc}
\tilde{\tilde{e}}_{k} &=& \tilde{\tilde{e}}_{\bar{k}} &=&  \tilde{e}_{k} - \left( \bar{V}^{p-h}_{knkn} \, - \, \bar{V}^{p-h}_{k\bar{n}k\bar{n}} \right) \, / \, 2  \,  \,  \,  \,  \, ,\\
 &&&& \\
&&&=&  \left(  \tilde{e}_{k} +  \tilde{e}_{\bar{k}}  \right) \, / \, 2  \,  \,  \,  \,  \, .\\
\label{diffenerind2}
\end{array}& 
\end{eqnarray}
\vspace{0.3cm}

The two successive energy differences can now
be given in terms of the full blocked state variables 
(variables defined with one tilde on top of them) using Eq.~\ref{diffener}:

\begin{eqnarray}
&\begin{array}{lcl}
\tilde{\tilde{E^{HFB}}} - E^{HFBE} &=& \frac{1}{2} \, \sum_{k \neq n, \bar{n}} \, \, \left(  e_{k} + \frac{\tilde{e}_{k} + \tilde{e}_{\bar{k}}}{2} \right) \, \left( \tilde{v}^{2}_{k} - v^{2}_{k} \right) \, + \left(  e_{n} + \frac{\tilde{e}_{n} + \tilde{e}_{\bar{n}}}{2} \right) \, \left( \frac{1}{2} -  v^{2}_{n} \right)   \,  \,  \,  \,  \, , \\
&& \\
&& - \, \frac{1}{2} \, \sum_{k} \left( \tilde{\Delta}_{k} \, \tilde{u}_{k} \, \tilde{v}_{k} \, - \, \Delta_{k} \, u_{k} \, v_{k}  \right) \\
\label{diffenercas1} \nb
\end{array}& 
\end{eqnarray}

\begin{eqnarray}
&\begin{array}{lcc}
\tilde{E^{HFB}} - E^{HFBE} &=& \frac{1}{2} \, \sum_{k} \, \left[  \, \left(  e_{k} + \tilde{e}_{k} \right) \, \left( \tilde{v}^{2}_{k} - v^{2}_{k} \right) \, - \, \tilde{\Delta}_{k} \, \tilde{u}_{k} \, \tilde{v}_{k} \, + \, \Delta_{k} \, u_{k} \, v_{k} \right]  \, \, \, \, \, \, \, \, . \\
\label{diffenercas2} \nb
\end{array}& 
\end{eqnarray}
\vspace{0.3cm}

We approximate the pure polarisation effect in the presence of 
pairing correlations 
by the energy difference $\tilde{E^{HFB}} - \tilde{\tilde{E^{HFB}}}$.
Using the last two equations, we obtain:

\begin{eqnarray}
E^{Pol} &=& \frac{\tilde{e}_{n}-\tilde{e}_{\bar{n}}}{4}  \, \, \, \, \, \, \, \, \,  \, \, , \\
\label{HFB_HFBE}
\end{eqnarray}

\noindent This result is formally identical to the HF result.

\end{appendix}

\section{Acknowledgment}
\label{secremer}

We thank the Department of Energy's Institute for Nuclear Theory at the University of Washington for its hospitality and partial support during the initialization of this work.

\clearpage


\newpage

\begin{footnotesize}

\btab
\caption[T2]{Nilsson numbers of created qp in odd cerium isotopes for zero as well as for realistic neutron pairing intensity (in MeV.fm$^{-3}$). Numbers are not reported when the created qp mixes several Nilsson states.}
\label{tabenergie}
\vspc
\begin{tabular}{rrrrrrrrrr}
\hh
   Cerium     &   &   $V_{n}^{p-p} = 0$ & & & & $V_{n}^{p-p} = 1250$  &  & \\
\hline
	&   &  & & K[Nn$_{z}$$\Lambda$] & &   &     &  \\
\hline
$^{119}$Ce &  &  3/2[422]        & & &&   3/2[422]    &  \\
$^{121}$Ce &  &  3/2[422]        & & &&   5/2[532]    &  \\
$^{123}$Ce &  &  1/2[541]        & & &&   5/2[413]    &  \\
$^{125}$Ce &  &  1/2[411]        & & &&   1/2[411]    &  \\
$^{127}$Ce &  &  7/2[523]        & & &&   5/2[402]    &  \\
$^{129}$Ce &  &  7/2[523]        & & &&   7/2[523]    &  \\
$^{131}$Ce &  &  7/2[404]        & & &&   7/2[404]    &  \\
$^{133}$Ce &  &  9/2[514]        & & &&               &  \\
$^{135}$Ce &  &  3/2[402]        & & &&               &  \\
$^{137}$Ce &  &  9/2[514]        & & &&               &  \\
$^{139}$Ce &  & 11/2[505]        & & &&               &  \\
$^{141}$Ce &  &  1/2[541]        & & &&   1/2[541]    &  \\
$^{143}$Ce &  &  3/2[532]        & & &&   3/2[532]    &  \\
$^{145}$Ce &  &  1/2[530]        & & &&   3/2[532]    &  \\
$^{147}$Ce &  &  1/2[660]        & & &&               &  \\
$^{149}$Ce &  &  3/2[651]        & & &&   3/2[521]    &  \\
$^{151}$Ce &  &  3/2[521]        & & &&   3/2[521]    &  \\
$^{153}$Ce &  &  5/2[642]        & & &&   3/2[521]    &  \\
$^{155}$Ce &  &  5/2[523]        & & &&   1/2[521]    &  \\
$^{157}$Ce &  &  1/2[521]        & & &&               &  \\
$^{159}$Ce &  &  5/2[512]        & & &&   5/2[512]    &  \\
$^{161}$Ce &  &  7/2[633]        & & &&   5/2[512]    &  \\
$^{163}$Ce &  &  1/2[600]        & & &&   7/2[633]    &  \\
$^{165}$Ce &  &  1/2[510]        & & &&   1/2[600]    &  \\
\hh   				
\etab

\end{footnotesize}
\end{document}